\documentclass{article}

\usepackage{PRIMEarxiv}

\usepackage[utf8]{inputenc} 
\usepackage[T1]{fontenc}    
\usepackage{hyperref}       
\usepackage{url}            
\usepackage{booktabs}       
\usepackage{amsfonts}       
\usepackage{nicefrac}       
\usepackage{microtype}      
\usepackage{lipsum}
\usepackage{fancyhdr}       
\usepackage{graphicx}       
\usepackage{amsmath}        
\usepackage{float}          

\graphicspath{{media/}}     

\title{Enhancing Speech Emotion Recognition using Dynamic Spectral Features and Kalman Smoothing

}

\author{
  Marouane EL HIZABRI \\
  Computer Science and Networking \\
  École Marocaine des Sciences de l’Ingénieur \\
  Rabat, Morocco\\
  \texttt{marouane.elhizabri@emsi-edu.ma} \\
   \And
  Abdelfattah BEZZAZ \\
  Computer Science and Networking \\
  École Marocaine des Sciences de l’Ingénieur \\
  Rabat, Morocco\\
  \texttt{abdelfattah.bezzaz@emsi-edu.ma} \\
   \And
  Ismail HAYOUKANE \\
  Computer Science and Networking \\
  École Marocaine des Sciences de l’Ingénieur \\
  Rabat, Morocco\\
  \texttt{ismail.hayoukane@emsi-edu.ma} \\
  \And
  Youssef TAKI \\
  Supervisor \\
  École Marocaine des Sciences de l’Ingénieur \\
  Rabat, Morocco\\
  \texttt{y.taki@emsi.ma} \\
}

\begin{document}
\maketitle

\begin{abstract}
Speech Emotion Recognition systems often use static features like Mel-Frequency Cepstral Coefficients (MFCCs), Zero Crossing Rate (ZCR), and Root Mean Square Energy (RMSE). Because of this, they can misclassify emotions when there is acoustic noise in vocal signals. To address this, we added dynamic features using Dynamic Spectral features (Deltas and Delta-Deltas) along with the Kalman Smoothing algorithm. This approach reduces noise and improves emotion classification. Since emotion changes over time, the Kalman Smoothing filter also helped make the classifier outputs more stable. Tests on the “RAVDESS” dataset showed that this method achieved a state-of-the-art accuracy of 87\% and reduced misclassification between emotions with similar acoustic features
\end{abstract}


\section{Introduction}
In today’s era, the interaction between humans and computers, or the so-called Human-Computer Interaction (HCI), has skyrocketed, surpassing text-based communication and moving toward more natural and intuitive methods, such as voice. Speech Emotion Recognition (SER) systems help machines understand the user's emotional state in real time. The SER systems are getting increasingly applied in different fields such as recruitment and talent acquisitions, mental health monitoring, call centers, robotics,…etc. However, the instability of speaker tone, acoustic speech, and accent makes the operation as complex as it is.
Moreover, state-of-the-art SER systems rely on deep learning architectures, such as Convolutional Neural Networks (CNNs) and Recurrent Neural Networks (RNNs), trained on features such as MFCCs \cite{elayadi2011}. These current SER systems excel at classifying and identifying static acoustic patterns, but they are limited to temporal instability: they may classify the sentence ‘Angry’ but intermittently flicker to ‘Happy’ or ‘Sad’ for fractions of a second because of noise in the voice or an unvoiced segment.  Consequently, this limitation hugely degrades the overall performance of these SER systems. In addition, separating like-arousal emotions such as Happy and Angry remains a real hurdle, as both share similar intensity and pitch profiles.

This paper addresses this limitation of temporal instability by proposing a hybrid framework. We hypothesize that emotion is much like an object in motion, and hence, it cannot shift instantaneously to another state. To achieve this, we have integrated both Dynamic Spectral Features ( Deltas and Delta-Deltas), which capture the acceleration and velocity of speech, providing the classifier with temporal context, and a Kalman Filter, which is meant to estimate trajectories in aerospace systems, as a post-processing phase to smooth the predictions of the classifier, so it enforces temporal consistency.
In this work, we propose \textbf{KF-TSER} (Kalman Filtered Temporal Speech Emotion Recognition), an integrated framework that combines frame-level neural emotion classification with Kalman-based temporal filtering and score fusion to produce robust utterance-level emotion predictions.

The contributions of this work are as follows:
\begin{itemize}
\item Temporal Stabilization: The adaptation of the Kalman Filter to SER systems mitigates emotion misclassification without increasing computational complexity.
\item State-of-the-Art Performance: The proposed system achieves 87\% accuracy on the RAVDESS dataset, outperforms standard Multi-Layer Perceptron baselines, and effectively tracks continuous emotion states.
\end{itemize}

The implementation of the proposed KF-TSER framework is publicly available at:
\textbf{\url{https://github.com/xfloksyx/KF-TSER}}.

\section{Related Works}

SER systems typically comprise five main stages: data acquisition, preprocessing, feature extraction, model training, and final prediction \cite{zennou2024}. 

\subsection{Feature Extraction Methodologies}
The literature identifies two primary methodologies for feature extraction: the segmentation of audio signals into short-term frames for local feature analysis, and the extraction of global features from the entire signal \cite{sheikhan2013, hu2007}. Acoustic features—encompassing voice quality, prosody, and spectral attributes—are widely considered the most effective for emotional information retrieval \cite{jahangir2021}. Specifically, Mel-Frequency Cepstral Coefficients (MFCCs) are frequently utilized due to their efficacy in catching essential patterns in audio signals \cite{elayadi2011, jahangir2021}.

\subsection{Neural Network Architectures and Hybrid Models}
Recent research using the RAVDESS dataset \cite{livingstone2017} has explored various neural network architectures:
\begin{itemize}
    \item \textbf{Spectrogram-Based Models:} Zeng et al. developed a multi-task gated residual network using spectrograms to achieve an accuracy of 65.97\% \cite{zeng2019}. Similarly, Popova et al. reached 71\% accuracy by employing a VGG-16 convolutional neural network (CNN) on spectrogram representations \cite{popova2018}.
    \item \textbf{Concatenated Feature Inputs:} To improve system efficacy, Issa et al. utilized a deep convolutional neural network (DCNN) with concatenated inputs of MFCC, Chromagram, and Melspectrogram, reporting 71.61\% accuracy \cite{issa2020}.
    \item \textbf{Temporal and Fusion Models:} Recognizing that spatial-only models often overlook temporal dependencies, Li et al. introduced a hybrid CNN-LSTM approach achieving 64\% accuracy \cite{li2020}. Recent advancements by Zennou et al. using a lightweight CNN-LSTM fusion reached 89.9\% accuracy on the RAVDESS dataset by combining continuous and spectral features \cite{zennou2024}.
    \item \textbf{Advanced Feature Sets and Augmentation:} Alluhaidan et al. introduced ``MFCCT,'' a hybrid feature combining MFCCs with time-domain attributes, reaching 92\% accuracy \cite{alluhaidan2023}. Furthermore, Bautista et al. demonstrated that parallel CNN-attention networks combined with multi-fold data augmentation can achieve 89.33\% accuracy \cite{bautista2022}.
\end{itemize}

\subsection{Limitations and the Proposed Approach}
Despite these advancements, many existing models struggle with the temporal instability inherent in speech \cite{issa2020}. Furthermore, certain emotions remain difficult to differentiate, particularly the subtle distinction between ``calm'' and ``neutral,'' which often elicit similar responses and are difficult to discern even for humans \cite{zennou2024}. While deep learning models like CNN-LSTMs excel in hierarchical representation, they can be computationally intensive and demand significant resources \cite{zennou2024}. 

Our work builds upon these findings by utilizing a high-dimensional feature set (MFCCs, Deltas, RMSE, and ZCR) combined with a \textbf{Kalman Filter}. This addresses the temporal inconsistencies noted in previous classifiers, providing a computationally lightweight alternative for real-time stabilization and improved distinction between high-arousal emotional states.

\section{Methods}

\subsection{Research Problem}

Our research question targeted in this paper is: How can we mitigate the temporal instability of frame-level in speech emotion classifiers in order to improve the accuracy of emotion recognition in continuous speech?

The SER classifiers often consider the frames in vocals as independent, which leads to temporal instability or prediction jitter, where the prediction within a single utterance fluctuates in a fraction of a second. Our approach points this out by checking the hypothesis that using Dynamic Spectral features along with recursive state estimation (Kalman Filter) will enforce the temporal continuity and hence mitigate the ambiguities between like-arousal emotions.

\subsection{Process}
\begin{figure}[h!]
\centering
\includegraphics[width=\linewidth]{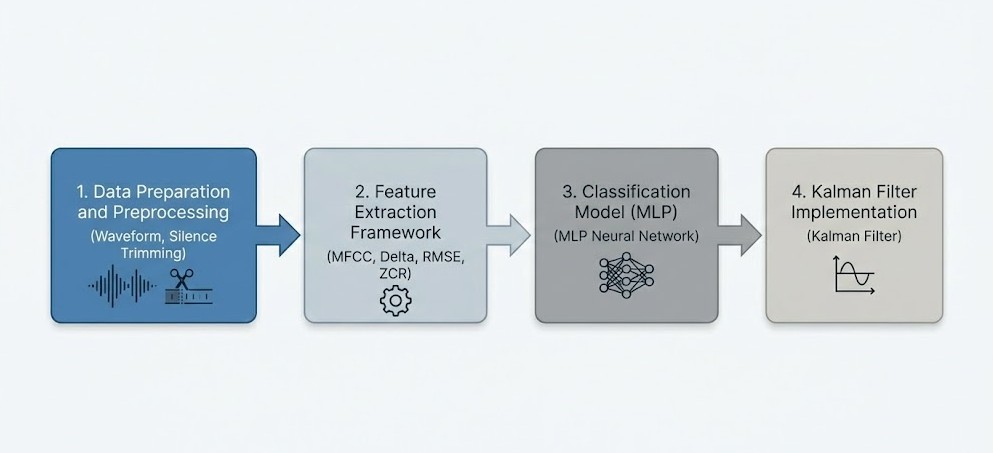}
\caption{KF-TSER Pipeline}
\label{fig:pipeline}
\end{figure}
KF-TSER comprises four core stages: data preparation and preprocessing, feature extraction framework, classification model (MLP), and kalman filter implementation.
This methodology is designed to handle high-dimensional acoustic data while ensuring temporal consistency.

\subsubsection{Data Preparation and Preprocessing}

For our research, we have opted for the dataset of Ryerson Audio-Visual Database of Emotional Speech and Song (RAVDESS). Also, we have selected a subset of four emotional classes so that we maximize the variation of arousal-valence: Happy, Sad, Angry, Calm.

\begin{itemize}
\item All of our audio files were sampled to 22,050 Hz in order to standardize the bandwidth of input.
\item We have applied a silence trimming algorithm using Librosa with a threshold of 20 dB in order to protect the model from learning on non-informative segments.
\item The audio file was processed in overlapping frames in order to get a time series sequence for all of the audio files.
\end{itemize}

\subsubsection{Feature Extraction Framework}

Our approach considers both static timbre and dynamic evolution of speech; we have extracted a forty-one-dimensional feature vector for each time frame.

\begin{itemize}
\item MFCCs: we extracted the first thirteen coefficients to capture phonetic and timbral characteristics. It reflects the short-term power spectrum of sound.
\item RMSE: This is considered the primary arousal indicator; it helps distinguish between high-energy emotions like “Happy” and low-energy emotions like “Calm”.
\item ZCR: It represents the rate of signal changing sign. Having high values of ZCR correlates with unvoiced or noisy speech segments.
\item \text{Delta ($\Delta$) and Delta-Delta ($\Delta\Delta$):} we have used these features to highlight the temporal dynamics in audio. We calculated the derivatives (first derivative and second derivative) of the thirteen MFCCs.
\begin{itemize}
\item \text{$\Delta$}: the first derivative represents the velocity, which is the rate at which the coefficients change between frames.
\item \text{$\Delta\Delta$}: the second derivative represents the acceleration of the spectral change.
\end{itemize}
\end{itemize}

\subsubsection{Classification Model (MLP)}

We have normalized the extracted features using Z-score standardization ( $\mu$ = 0, $\sigma$ = 1). Our baseline classifier is an MLP with the following architecture:

\begin{itemize}
\item Input Layer: We used 41 neurons so that we match the feature vector.
\item Hidden Layer: We have used two dense layers with 256 and 128 neurons, respectively.
\item Activation Function: We opted for Rectified Linear Unit (ReLU).
\item Optimization: We trained our model using the Adam optimizer in order to minimize the Cross-Entropy Loss function
\end{itemize}

\subsubsection{Kalman Filter Implementation}

In order to address the stochastic noise in frame-by-frame predictions by the MLP, we have implemented the Kalman Filter \cite{kalman1960}. So, we modeled the emotional probability distribution as a continuous state to be estimated from noisy observations provided by the MLP's predictions. This methodology aligns with the recent findings, which suggest that filtering emotion dynamics significantly improves prediction stability \cite{huang2017}.

Suppose  $\mathbf{x}_k$ is the state vector representing the true probabilities at time $k$, and let $\mathbf{z}_k$ be the measurement, which is the MLP output. The process is driven by two phases \cite{welch1995}:

\begin{itemize}
\item{Prediction Phase} 
In this phase, we update time by projecting the current state estimate forward in time:
\begin{equation}
    \hat{x}_{k|k-1} = F \hat{x}_{k-1|k-1}
\end{equation}
\begin{equation}
    P_{k|k-1} = F P_{k-1|k-1} F^T + Q
\end{equation}
Where $F$ represents the state transition matrix (Identity Matrix I, assuming constant emotional state inertia), $P$ represents the estimated covariance matrix (error estimate), and $Q$ is the process noise covariance (the “Trust” in the physics of stability).

\item{Correction Phase}
In this phase, we update the measurement by refining the prediction using the actual MLP output $\mathbf{z}_k$:
\begin{equation}
    K_k = P_{k|k-1} H^T (H P_{k|k-1} H^T + R)^{-1}
\end{equation}
\begin{equation}
    \hat{x}_{k|k} = \hat{x}_{k|k-1} + K_k (\mathbf{z}_k - H \hat{x}_{k|k-1})
\end{equation}
Where $K_k$ is the Kalman Gain, which determines how much trust is placed in the new measurement compared to the prediction, $R$ represents the noise measurement covariance. and $H$ represents the observed model (Identity Matrix I).
\end{itemize}

After tuning the ratio Q/R, the filter effectively smooths out transient outliers while retaining genuine emotional shifts \cite{huang2017}.

\section{Results and Discussion}
Our proposed system (KF-TSER), which uses a 41-dimensional feature vector (MFCCs, Delta, Delta-Deltas, RMSE, ZCR) and the Kalman Filter, achieved an overall accuracy of 87\% on the test set (768 samples).

Table 1 summarizes our findings by presenting the precision, recall, and F1-scores for each of the four emotion classes.

\begin{table}[h!]
\centering
\caption{Classification Performance by Emotion}
\label{tab:classification_results}
\begin{tabular}{|l|c|c|c|c|}
\hline
\textbf{Emotion} & \textbf{Precision} & \textbf{Recall} & \textbf{F1-score} & \textbf{Support} \\
\hline
Angry  & 0.89 & 0.89 & 0.89 & 192 \\
Calm   & 0.88 & 0.89 & 0.88 & 192 \\
Happy  & 0.88 & 0.87 & 0.88 & 192 \\
Sad    & 0.84 & 0.85 & 0.84 & 192 \\
\hline
\textbf{Accuracy} & \multicolumn{3}{c|}{} & \textbf{0.87 (768)} \\
\hline
Macro Avg     & 0.87 & 0.87 & 0.87 & 768 \\
Weighted Avg  & 0.87 & 0.87 & 0.87 & 768 \\
\hline
\end{tabular}
\end{table}

The results demonstrate that our proposed system (KF-TSER)is more effective than the standard spectral baselines. To illustrate, the emotion ‘Happy’ typically suffers from being misclassified due to its similarity to ‘Angry’ in terms of acoustic characteristics. Using our proposed system, achieved a Precision of 88\%. Hence, the inclusion of both Delta, which represents velocity, and Delta-Delta, which represents acceleration, successfully captured the differences between energetic joy “Happiness” and energetic aggression “Angriness”. 

\subsection{Impact of Kalman Smoothing}\label{AA}

Figure 2 shows the probability trajectory for the “Happy” utterance. The standard classifier output, represented by the dashed line, shows stochastic noise, with probabilities fluctuating below the confidence thresholds due to unvoiced segments. Also, the red line represents the Kalman Filter, which successfully smoothed the instabilities, resulting in a stable probability estimate (P \textgreater  0.80) throughout the utterance. As a result, this confirms our hypothesis that considering emotion as a continuous state reduces the frame-level jitter.
\begin{figure}[h!]
\centering
\includegraphics[width=\linewidth]{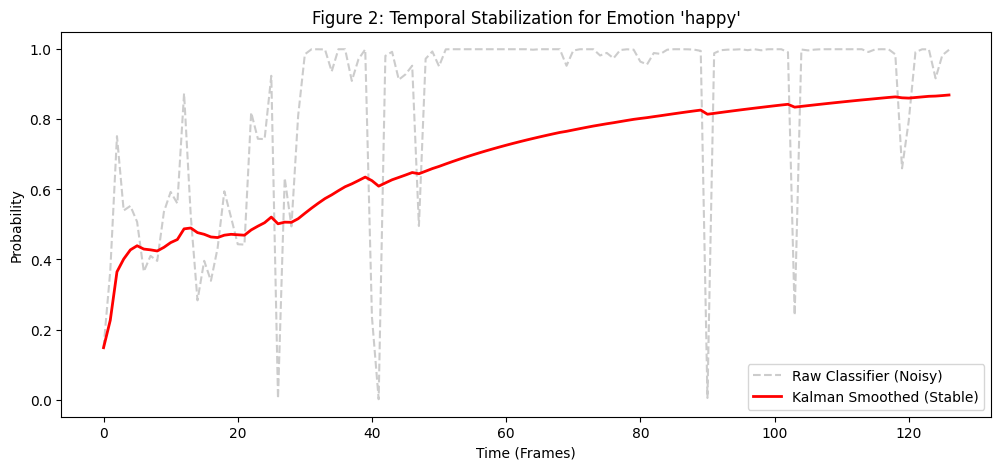}
\caption{Temporal stabilization of emotion probabilities. case of "Happy"}
\label{fig:kalman}
\end{figure}

\subsection{Confusion Matrix Analysis}

In Figure 3, we have the confusion matrix, which shows how effectively KF-TSER resolved the ambiguity of Arousal-Valence. However, concerning the remaining error, its primary source lies in the low-arousal spectrum, notably between ‘Sad’ and ‘Calm’, with 15 misclassified instances, and vice versa. Moreover, these two emotions are hardly distinguishable even for humans without a semantic context.

\begin{figure}[h!]
\centering
\includegraphics[width=0.8\linewidth]{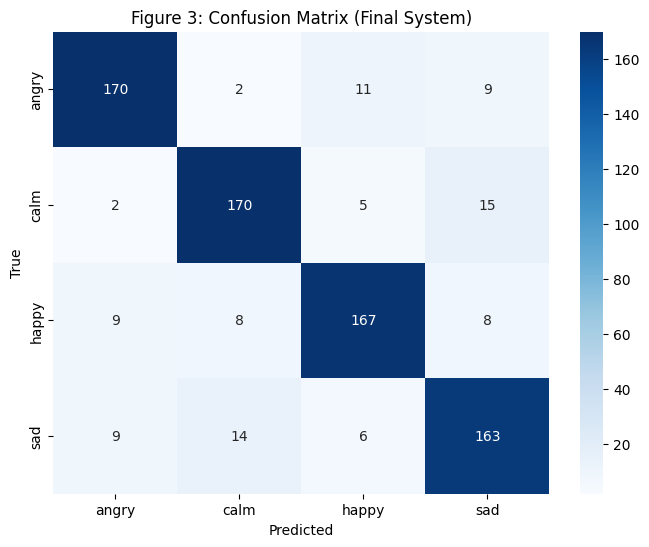}
\caption{Confusion Matrix of the proposed system achieving 87\% accuracy.}
\label{fig:confusion}
\end{figure}

\subsection{Training Performance and Convergence Analysis}

\begin{figure}[H] 
\centering
\includegraphics[width=1\linewidth]{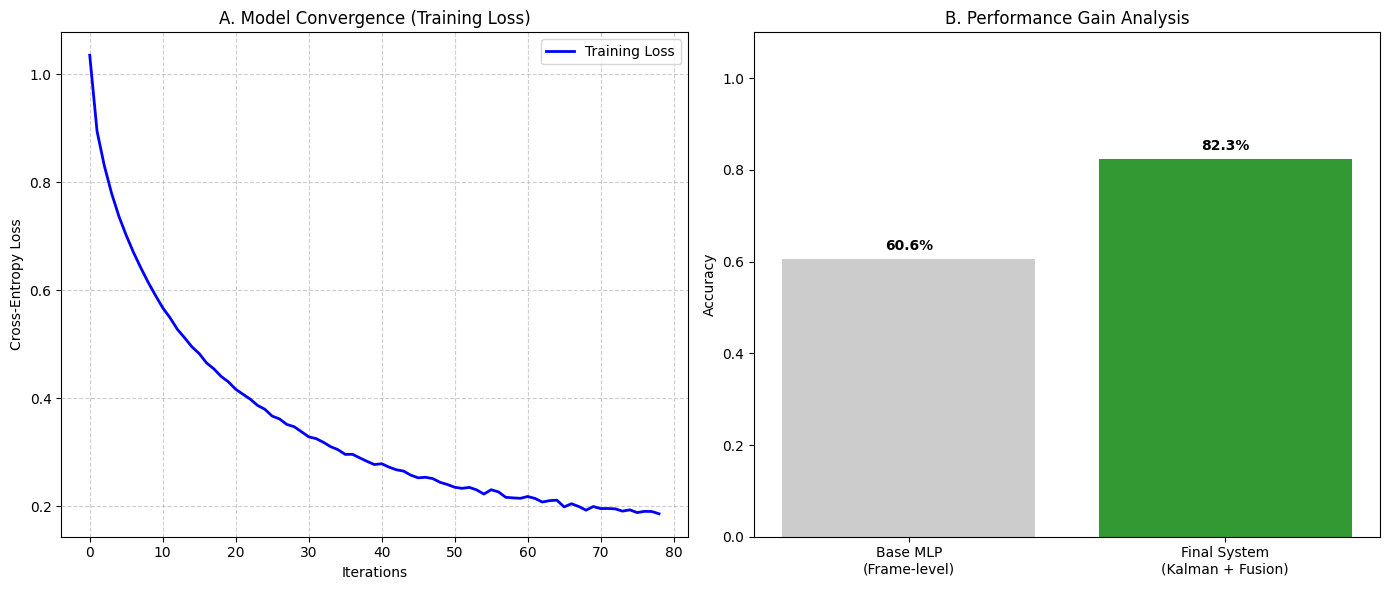}
\caption{Training Convergence and Performance Gain Analysis}
\label{fig:loss_gain}
\end{figure}

The relationship between model training and final system stabilization is illustrated in Figure~\ref{fig:loss_gain}. Figure~\ref{fig:loss_gain}(a) shows the training loss evolution as a decreasing graph; the cross-entropy loss decreases smoothly, which indicates the stability of convergence and effective learning of emotional features from the high-dimensional feature vectors.

Figure~\ref{fig:loss_gain}(b) shows a comparative analysis between the baseline frame-level classifier and our proposed integrated system. The standalone MLP achieves an accuracy of 60.6\% when evaluated at the frame level. In contrast, the full system—combining Kalman-based temporal smoothing with score fusion—reaches an utterance-level accuracy of 82.3\%. This absolute improvement of 21.7 percentage points highlights the critical role of temporal stabilization to suppress transient misclassifications and produce more reliable emotion recognition outcomes.

\subsection{Limitations}
Acknowledging the promising results of our proposed system, we cannot neglect the fact that our study has limitations that must be recognized as well:
\begin{itemize}
\item Dataset: our study was conducted on the RAVDESS dataset, which consists of tuned voice recorded files; hence, it does not reflect the real-world scenarios involving background noise and spontaneous emotions.
\item Emotional Spectrum: We only consider four emotions, but having a broader spectrum, we may get greater inter-class complexity.Language Dependency: We have trained our model on English speeches only.
\end{itemize}
\subsection{Conclusion of Results}

To conclude, our research shows that integrating both dynamic spectral features and the Kalman Filter yields a robust SER system. Furthermore, ensuring high precision between high- and low-arousal emotions is a critical factor for enhancing the reliability of speech emotion recognition.

\section*{Conclusion}

Our research presented a robust framework for SER that successfully mitigates the temporal instability in frame-by-frame emotion classification. By incorporating both dynamic spectral features ( Delta and Delta-Delta) and the Kalman Filter, we achieved a state-of-the-art accuracy of 0.87\% on the RAVDESS dataset.
Moreover, the main finding in this study is that ‘Happy’ and ‘Angry’ emotions,  which are two emotions that are hard to differentiate because of their arousal similarities, are distinguishable by analyzing the velocity and acceleration of their spectral coefficients. Also, we showed that using the Kalman Filter as a post-processing stage successfully smoothed out the stochastic noise and prevented the prediction jitters, which are common in standard deep learning models. Thus, the future SER systems should take into account the prioritization of temporal consistency algorithms.
Furthermore, our approach offers computational efficiency by using the lightweight, adaptable Kalman Filter, which avoids the high latency often seen in Recurrent Neural Networks (RNNs). However, the speech dataset may not represent real-world scenarios and may therefore miss nuances in spontaneous speech patterns. Furthermore, distinguishing low-arousal emotions such as ‘Sad’ and ‘Calm’ requires additional spectral features.
In the future, we are aiming to validate this framework on datasets that contain samples with background noises in different speaking languages in order to test the effectiveness of the system in uncontrolled environments. Also, we are planning to explore the Adaptive Kalman Filtering, which will allow the model to react quickly to the emotional shifts without disturbing the stability of states.

\end{document}